# 18

# Cellular Automata, PDEs, and Pattern Formation

Xin-She Yang and
Y. Young

## 18.1 Introduction

A cellular automaton (CA) is a rule-based computing machine, which was first proposed by von Newmann in early 1950s and systematic studies were pioneered by Wolfram in 1980s. Since a cellular automaton consists of space and time, it is essentially equivalent to a dynamical system that is discrete in both space and time. The evolution of such a discrete system is governed by certain updating rules rather than differential equations. Although the updating rules can take many different forms, most common cellular automata use relatively simple rules. On the other hand, an equation-based system such as the system of differential equations and partial differential equations also describe the temporal evolution in the domain. Usually, differential equations can also take different forms that describe various systems. Now one natural question is what is the relationship between a rule-based system and an equation-based system?

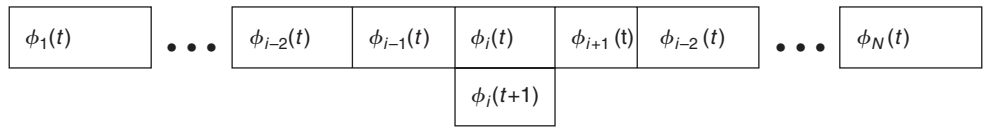

**FIGURE 18.1**   Diagram of a one-dimensional cellular automaton.

Given differential equations, how can one construct a rule-based cellular automaton, or vice versa? There has been substantial amount of research in these areas in the past two decades. This chapter intends to summarize the results of the relationship among the cellular automata, partial differential equations (PDEs), and pattern formations.

## 18.2  Cellular Automata

### 18.2.1  Fundamentals of Cellular Automaton

On a one-dimensional (1D) grid that consists of $N$ consecutive cells, each cell $i$ ($i = 1, 2, \ldots, N$) may be at any of the finite number of states, $k$. At each time step, $t$, the next state of a cell is determined by its present state and the states of its local neighbors. Generally speaking, the state $\phi_i$ at $t + 1$ is a function of its $2r + 1$ neighbors with $r$ cells on the left of the concerned cell and $r$ cells on its right (see Figure 18.1). The parameter $r$ is often referred to as the radius of the neighborhood.

The number of possible permutations for $k$ finite states with a radius of $r$ is $p = k^{2r+1}$, thus the number of all possible rules to generate the state of cells at next time step is $k^p$, which is usually very large. For example, if $r = 2$, $k = 5$, then $p = 125$ and the number of possible rules is $5^{125} \approx 2.35 \times 10^{87}$, which is much larger than the number of stars in the whole universe.

The rule determining the new state is often referred to as the transition rule or updating rule. In principle, the state of a cell at next time step can be any function of the states of some neighbor cells, and the function can be linear and nonlinear. There is a subclass of the possible rules according to which the new state depends only on the sum of the states in a neighborhood, and this will simplify the rules significantly. For this type of updating rules, the number of possible permutations for $k$ states and $2r + 1$ neighbors is simply $k(2r + 1)$, and thus the number of all possible rules is $k^{k(2r+1)}$. Then for the same parameter $k = 5$, $r = 2$, the number of possible rules is $5^{15} = 3.05 \times 10^{10}$, which is much smaller when compared with $5^{125}$. This subclass of sum-rule or totalistic rule is especially important in the popular cellular automata such as Conway's Game of Life, and in the cellular automaton implementation of partial differential equations.

The dynamics and complexity of cellular automata are extremely rich. Stephan Wolfram's pioneering research and mathematical analysis of cellular automata led to his famous classification of 1D cellular automata:

*Class 1*: The first class of cellular automata always evolves after a finite number of steps from almost all initial states to a homogeneous state where every cell is in the same state. This is something like fixed point equilibrium in the dynamical system.

*Class 2*: Periodic structures with a fixed number of states occur in the second class of cellular automata.

*Class 3*: Aperiodic or "chaotic" structures appear from almost all possible initial states in this type of cellular automata.

*Class 4*: Complex patterns with localized spatial structure propagate in the space as time evolves. Eventually, these patterns will evolve to either homogeneous or periodic. It is suggested that this class of cellular automata may be capable of universal computation.

Cellular automata can be formulated in higher dimensions such as 2D and 3D. One of the most popular and yet very interesting 2D cellular automata using relatively simple updating rules is the Conway's Game of Life. Each cell has only two states $k = 2$, and the states can be 0 and 1. With a radius of $r = 1$ in the 2D case, each cell has eight neighbors, thus the new state of each cell depends on total nine cells surrounding it. The boundary cells are treated as periodic. The updating rules are: if two or three neighbors of a cell are alive (or 1) and it is currently alive, then it is alive at next time step; if three neighbors of a cell are alive (or 0) and it is currently not alive its next state is alive; the next state is not alive for all the other cases. It is suggested that this simple automaton may have the capability of universal computation. There are many existing computer programs such as Life in Matlab and screen savers on all computer platforms such as Windows and Unix.

## 18.2.2 Finite-State Cellular Automata

In general, we can define a finite-state cellular automaton with a transition rule $G = [g_{ij,...,l}]$, $(i, j, ..., l = 1, 2, ..., N)$ from one state $\Phi^t = [\phi^t_{ij,...,l}]$ at time level $n$ to a new state $\Phi^{t+1} = [\phi^{t+1}_{ij,...,l}]$ at a new time step $n + 1$. The value of subscript $(i, j, ..., l)$ denotes the dimension, $d$, of the cellular automaton. Therefore, a CA in the $d$-dimensional space has $N^d$ cells. For the 2D case, this can be written as

$$G : \Phi^t \mapsto \Phi^{t+1}, \quad g_{ij} : \phi^t_{ij} \mapsto \phi^{t+1}_{ij}, \quad (i, j = 1, 2, ..., N).$$

In the case of sum-rule with a $4r + 1$ neighbors, this becomes

$$\phi^{t+1}_{ij} = G\left( \sum_{\alpha=-r}^{r} \sum_{\beta=-r}^{r} a_{\alpha\beta} \phi^t_{i+\alpha, j+\beta} \right), \quad (i, j = 1, 2, ..., N),$$

where $a_{\alpha\beta}$ $(\alpha, \beta = \pm 1, \pm 2, ..., \pm r)$ are the coefficients. The cellular automata with fixed rules defined this way are deterministic cellular automata. In contrast, there exists another type, namely, the stochastic cellular automata that arise naturally from the stochastic models for natural systems (Guinot, 2002; Yang, 2003).

## 18.2.3 Stochastic Cellular Automata

When using cellular automata to simulate the phenomena with stochastic components or noise such as percolation and stochastic process, the more effective way is to introduce some probability associated with certain rules. Usually, there is a set of rules and each rule is applied with a probability (Guinot, 2002). Another way is that the state of a cell is updated according to a rule only if certain conditions are met or certain values are reached for some random variables. For example, the rule for 2D a cellular automaton $g(\phi^t_{ij}) = \phi^{t+1}_{ij}$ is applied at a cell only if a random variable $\nu \leq \Gamma(\phi^t_{ij})$ where the function $\Gamma \in [0, 1]$. At each time step, a random number $\nu$ is generated for each cell $(i, j)$, and the new state will be updated only if the generated random number is greater than $\Gamma$, otherwise, it remains unchanged. Cellular automata constructed this way are called stochastic or probabilistic cellular automata. An example is given later in the next section.

## 18.2.4 Reversible Cellular Automata

A cellular automaton with an updating rule $\phi^{t+}_{ij} = g(\phi^t_{ij})$ is generally irreversible in the sense that it is impossible to know the states of a region such as all zeros were the same at a previous time step or not. However, certain class of rules will enable the automata to be reversible. For example, a simple finite difference (FD) scheme for a dynamical system

$$u(t + 1) = g[u(t)] - u(t - 1) \quad \text{or} \quad u(t - 1) = g[u(t)] - u(t + 1),$$

is reversible since for any function $g(u)$, one can compute $u(t + 1)$ from $u(t)$ and $u(t - 1)$, and invert $u(t - 1)$ from $u(t)$ and $u(t + 1)$. The automaton rule for 2D reversible automata can be similarly constructed as

$$u_{i,j}^{t+1} = g(u_{i,j}^{t}) - u_{i,j}^{t-1},$$

together with appropriate boundary conditions such as fixed-state boundary conditions (Margolus, 1984).

## 18.3   Cellular Automata for PDEs

Cellular automata have been used to study many phenomena (Vichniac, 1984; Wolfram, 1984, 1994; Weimar, 1997; Yang, 2002). In fact, many natural phenomena behave like finite-state cellular automata, and these include self-organized systems such as pattern formation in biological system, insect colonies, and ecosystem; multiple particle system such as the lattice-gas, granular material, and fluids; autocatalytic systems such as enzyme functionality and mineral reactions; and even systems involving society and culture interactions. However, most of these systems have been studied using continuum-based differential equations. We now focus on the formulation of automaton rules from corresponding PDEs.

### 18.3.1   Rules-Based System and Equation-Based System

Equation-based relationships, often in the form of ordinary differential equations and partial differential equations, form the continuum models of most physical, chemical, and biological processes. Differential equations are suitable and work well for systems with only a small degree of freedom and evolution of system variables in the continuous and smooth manner. There have been vast literatures on analysis and solution technique of the differential models. On the other hand, cellular automata are often considered as an alternative approach and may compliment the existing mathematical basis. The state variables are always discrete, but the numbers of degree of freedom are large (Wolfram, 1984).

Although continuum models have advantages such as high accuracy and conservation laws, mathematical analysis are usually very difficult and the analytical solutions do not always exist. Only few differential equations have a closed-form solution. In last several decades, the numerical methods have become the essential parts of the solution and of the analysis of almost all problems in engineering, physics, and biology. In fact, computational modeling and numerical computation have become the third component, bridging the gap between theoretical models and experiments. As the computing speed and memory of computers increase, computer simulations have become a daily routine in science and engineering, especially in multidisciplinary research. There are also vast literatures concerning the numerical algorithms, numerical solutions of partial differential equations, and others. These include the following well-established methods: finite difference method, finite element method, finite volume method, cellular automata, lattice-gas, Monte-Carlo method, and genetic algorithms.

Finite difference methods work very well for many problems, but have disadvantages in dealing with irregular geometry. Finite element and finite volume methods can deal with irregular geometry and thus are commonly used and there are quite a few commercial software packages available. Lattice-gas and Monte-Carlo methods are suitable for many problems in physics, especially for systems with multibodies or multiparticles. Cellular automaton is a very interesting and powerful method, and it has gradually become an essential part of the numerical computations owing to its universal computability and the nature of parallel implementation.

### 18.3.2   Finite Difference Scheme and Cellular Automata

There is a similarity between finite difference scheme and finite-state cellular automata. Finite difference scheme is the discretization of a differential equation on a regular grid of points with the evolution of the states over the discrete time steps. Even the state variable from a differential equation may have continuous values; numerical computation on a computer always lead to the discrete values due to the limited bits of processors or round-off. Similarly, cellular automata are also about the evolution of state variables with finite number of values on a regular grid of cells at different discrete time steps. If the number of states of a cellular automaton is comparable with that of the related finite difference equation, then we can expect the results to be comparable.

To demonstrate this, we choose the 1D heat equation

$$\frac{\partial T}{\partial t} = \kappa \frac{\partial^2 T}{\partial x^2},$$

where $T$ is temperature and $\kappa$ is the thermal diffusivity. This is a mathematical model that is widely used to simulate many phenomena. The temperature $T(x, t)$ is a real-valued function, and it is continuous for any time $t > 0$ whatever the initial conditions. In reality, it is impossible to measure the temperature at a mathematical point, and the temperature is always the average temperature in a finite representative volume over certain short time. Mathematically, one can obtain a closed-form solution with infinite accuracy in the domain, but physically the temperature would only be meaningful at certain macroscopic levels. No matter how accurate the solution may have at very fine scale, it would be meaningless to try to use the solution at the atomic or subatomic levels where quantum mechanics come into play and, the solution of temperature is invalid (Toffoli, 1984). Thus, numerical computation would be very useful even though it has finite discrete values.

The simplest discretization of the above heat equation is the central difference for spatial derivative and forward scheme for time derivatives, and we have

$$T_i^{n+1} - T_i^n = \frac{\kappa \Delta t}{(\Delta x)^2}(T_{i+1}^n - 2T_i^n + T_{i-1}^n),$$

where $i$ and $n$ are the spatial and time indices. If we choose the time steps and spatial discretization such that $\kappa \Delta t/(\Delta x)^2 = 1$, now we have

$$T_i^{t+1} = (T_{i+1}^t + T_i^t + T_{i-1}^t) - 2T_i^t,$$

which is something like the "mod 2" cellular automata or Wolfram's cellular automata with rule 150 (Wolfram, 1984; Weimar, 1997).

Cellular automata obtained this way are very similar to the finite difference method. If the state variables are discrete, they are exactly the finite-state cellular automata. However, one can use the continuous-valued state variable for such simulation, in this case, they are continuous-valued cellular automata from differential equations as studied by Rucker's group and used in their the well-known CAPOW program (Rucker et al., 1998; Ostrov and Rucker, 1997). In some sense, the continuous-valued cellular automata based on the differential equations are the same as the finite difference methods, but there are some subtle differences and advantages of cellular automata over the finite difference simulations due to the CA's properties of parallel nature, artificial life-oriented emphasis on experiment and observation and genetic algorithms for searching large phase space of the rules as proposed by Rucker et al. (1998).

### 18.3.3   Cellular Automata for Reaction-Diffusion Systems

By extending the discretization procedure from differential equations to derive automaton rules for cellular automata, we now formulate the cellular automata from their corresponding partial differential equations. First, let us start with the reaction-diffusion equation that may form beautiful patterns in the 2D configuration

$$\frac{\partial u}{\partial t} = D\left(\frac{\partial^2 u}{\partial x^2} + \frac{\partial^2 u}{\partial y^2}\right) + f(u),$$

where $u(x, y, t)$ is the state variable that evolves with time in a 2D domain, and the function $f(u)$ can be either linear or nonlinear. $D$ is a constant depending on the properties of diffusion. This equation can also be considered as a vector form for a system of reaction-diffusion equations if let $D = \text{diag}(D_1, D_2)$, $u = [u_1 u_2]^T$. The discretization of this equation can be written as

$$\frac{u_{i,j}^{n+1} - u_{i,j}^n}{\Delta t} = D \left[ \frac{u_{i+1,j}^n - 2u_{i,j}^n + u_{i-1,j}^n}{(\Delta x)^2} + \frac{u_{i,j+1}^n - 2u_{i,j}^n + u_{i,j-1}^n}{(\Delta y)^2} \right] + f(u_{i,j}^n),$$

by choosing $\Delta t = \Delta x = \Delta y = 1$, we have

$$u_{i,j}^{n+1} = D[u_{i+1,j}^n + u_{i-1,j}^n + u_{i,j+1}^n + u_{i,j-1}^n] + f(u_{i,j}^n) + (1 - 4D)u_{i,j}^n,$$

which can be written as the generic form

$$u_{i,j}^{t+1} = \sum_{k,l=-r}^{r} a_{k,l} u_{i+k,j+l}^t + f(u_{i,j}^t),$$

where the summation is over the $4r + 1$ neighborhood. This is a finite-state cellular automaton with the coefficients $a_{k,l}$ being determined from the discretization of the governing equations, and for this special case, we have $a_{-1,0} = a_{+1,0} = a_{0,-1} = a_{0,+1} = D$, $a_{0,0} = 1 - 4D$, $r = 1$.

### 18.3.4 Cellular Automata for the Wave Equation

For the 1D linear wave equation,

$$\frac{\partial^2 u}{\partial t^2} = c^2 \frac{\partial^2 u}{\partial x^2},$$

where $c$ is the wave speed. The simplest central difference scheme leads to

$$\frac{u_i^{n+1} - 2u_i^n + u_i^{n-1}}{(\Delta t)^2} = c^2 \frac{u_{i+1}^n - 2u_i^n + u_{i-1}^n}{(\Delta x)^2}.$$

By choosing $\Delta t = \Delta x = 1$, $t = n$, it becomes

$$u_i^{t+1} = [u_{i+1}^t + u_{i-1}^t + 2(1 - c^2)u_i^t] - u_i^{t-1}.$$

This can be written in the generic form

$$u_i^{t+1} + u_i^{t-1} = g(u^t),$$

which is reversible under certain conditions. This property comes from the reversibility of the wave equation because it is invariant under the transformation: $t \rightarrow -t$.

### 18.3.5 Cellular Automata for Burgers Equation with Noise

One of the important equations arising in many processes such as turbulent phenomenon is the noisy Burgers equation

$$\frac{\partial u}{\partial t} = 2u \frac{\partial u}{\partial x} + \frac{\partial^2 u}{\partial x^2} + \nabla \upsilon,$$

where $\upsilon$ is the noise that is uncorrelated in space and time so that $\langle \upsilon(x,t) \rangle = 0$ and $\langle \upsilon(x,t)\upsilon(x_0,t_0) \rangle = 2D\delta(x - x_0)\delta(t - t_0)$ (Emmerich and Hahng, 1998). This equation with Gaussian white noise can be rewritten as

$$\frac{\partial u}{\partial t} + \xi = 2u\frac{\partial u}{\partial x} + \frac{\partial^2 u}{\partial x^2} + \eta,$$

where both $\xi$ and $\eta$ are uncorrelated. By introducing the variables $v_i^t = c \exp(\Delta x\, u_i^t), \phi_i^t = \beta \ln(v_i^t), \alpha = \Delta t/(\Delta x)^2, (1 - 2\alpha)/c\alpha = \exp(-A/\beta), c^2 = \exp(B/\beta), \xi = \exp(\Phi), \eta = \exp(\Psi)$ and after some straightforward calculations in the limit of $\beta$ tends zero, we have the automata rule

$$\phi_i^{t+1} = \phi_{i-1}^t + \max[0, \phi_i^t - A, \phi_i^t + \phi_{i+1}^t - B, \Psi_i^t - \phi_{i-1}^t]$$
$$- \max[0, \phi_{i-1}^t - A, \phi_{i-1}^t + \phi_i^t - B, \Phi_i^t - \phi_{i-1}^t],$$

where we have used the following identity,

$$\lim_{\phi \to +0} \varepsilon \ln(e^{A/\varepsilon} + e^{B/\varepsilon} + \cdots) = \max[A, B, \ldots].$$

This forms a generalized probabilistic cellular automaton that is referred to as the noisy Burgers cellular automaton. Burgers equation without noise usually evolves in shock wave, and in the presence of noise, the states of the probabilistic cellular automata may be taken as discrete reminders of those shock waves that were disorganized.

## 18.4   Differential Equations for Cellular Automata

Computer simulations based on cellular automata and partial differential equations work remarkably well for many different reasons. One possibility is that finite-state cellular automata and finite difference approximations using discrete time and space with a finite precision can represent physical variables very well and thus the models are insensitive to very small space and time scales. It is relatively straight-forward to derive the updating rules for cellular automata from the corresponding partial differential equations. However, the reverse is usually very difficult. There is no general method available to formulate the continuum model or differential equations for given rule-based cellular automata despite the obvious importance. Fortunately, there have been some important progress made in this area (Omohundro, 1984; Wolfram, 1984), and we outline some of the procedures of formulating continuum equations for cellular automata.

### 18.4.1   Formulation of Differential Equations from Cellular Automata

The mathematical analysis for cellular automata was first pioneered by Wolfram, and has attracted wide attention since 1980s. Wolfram (1983) gave an extensive analysis of statistical mechanics of cellular automata. Later on, Omohundro (1984) provided an instructive procedure of formulating the partial differential equations for cellular automata by using 10 PDE variables in 2D configurations. These ten variables are the state variable $P(x, y, t)$ at present, new state $N(x, y, t)$ and eight variables $U_1, \ldots, U_8$ with eight bump or bell functions. The variables have the same format of information as the $P(x, y, t)$. On a 2D grid, these eight functions $S_1, \ldots, S_8$ are shown in Figure 18.2 together with shifted coordinates.

According to Omohundro's formulation, we assume the bumps to be $\alpha$ wide and constant outside $\beta$ from the transition. If the width of a cell is 1, then $\alpha = 1/5$ and $\beta = 1/100$. The eight functions are

| $S_2$ (−1,+1) | $S_3$ (0,+1) | $S_4$ (+1,+1) |
|---|---|---|
| $S_1$ | N/P | $S_5$ |
| $S_8$ (−1,−1) | $S_7$ (0,−1) | $S_6$ (+1,−1) |

**FIGURE 18.2**   Diagram of eight neighbors and the positions of Omohundro's functions.

taken as

$$S_1 = e^{-1/(\beta-x)^2} \; (-\beta < x < \beta), \; 0 \; (|x| \geq \beta), \quad S_2 = \frac{S_1(x\beta/\alpha)}{S_1(0)}, \quad S_3 = \int_{-1}^{x} S_1(x)\mathrm{d}x \Big/ \int_{-\beta}^{\beta} S_1(x)\mathrm{d}x,$$

$$S_4 = S_3(x - \alpha/2)S_3(\alpha/2 - x), \quad S_5 = S_3(x - \alpha)S_3(\alpha - x), \quad S_6 = \sum_{k=-\infty}^{\infty} S_4(x-k),$$

$$S_7 = \sum_{k=-\infty}^{\infty} S_5(x-k), \quad S_8 = \sum_{k=-\infty}^{\infty} S_2(x-k).$$

By using these functions, Omohundro derived the following equation:

$$\frac{\partial N}{\partial t} = -\gamma \left[ \frac{\mathrm{d}S_2}{\mathrm{d}x}\frac{\partial N}{\partial x} + \frac{\mathrm{d}S_2}{\mathrm{d}y}\frac{\partial N}{\partial y} \right],$$

where $\gamma$ is a large constant. Differential equations for other variables can be written in a similar manner although they are more complicated (Omohundro, 1984).

### 18.4.2   Stochastic Reaction-Diffusion

A stochastic cellular automaton on a 1D ring has $N$ sites with simple rules for the state variable $u_i(t)$ and $S = u_i(t) + u_{i+1}(t)$: (1) $u_i(t+1) = 0$, if $S = 0$; (2) $u_i(t+1) = 1$ with probability $p_1$, if $S = 1$; (3) $u_i(t+1) = 1$ with probability $p_2$, if $S = 2$. This Domany–Kinzel cellular automaton is equivalent to the following stochastic reaction-diffusion equation:

$$\frac{\partial v}{\partial t} = D\nabla^2 v + f(v) + \varepsilon\sqrt{v},$$

where $v$ is the concentration of live sites or $u_i(t) = 1$, and $\varepsilon$ is a zero-mean Gaussian random variable with unit variance (Ahmed and Elgazzar, 2001). However, there is nonuniqueness associated with the formulation of differential equations from the cellular automata. Bognoli et al. (2002) demonstrated that the above equation can be obtained from the following rules: (1) $u_i(t+1) = 0$, if $\Sigma = u_{i+1}(t) + u_i(t) + u_{i-1}(t) = 0$; (2) $u_i(t+1) = 1$ with probability $p_1$, if $\Sigma = 1$; (3) $u_i(t+1) = 2$ with probability $p_2$, if $\Sigma = 2$; (4) $u_i(t+1) = 1$, if $\Sigma = 3$. This nonuniqueness in the relationship between cellular automata and PDEs require more research.

## 18.5   Pattern Formation

The behavior and characteristics of cellular automata are very complicated and there is no general or universal mathematical analysis available for the description of such complexity. Even for 1D cellular automata, they are complicated enough as shown by Wolfram classifications. Pattern formation is one

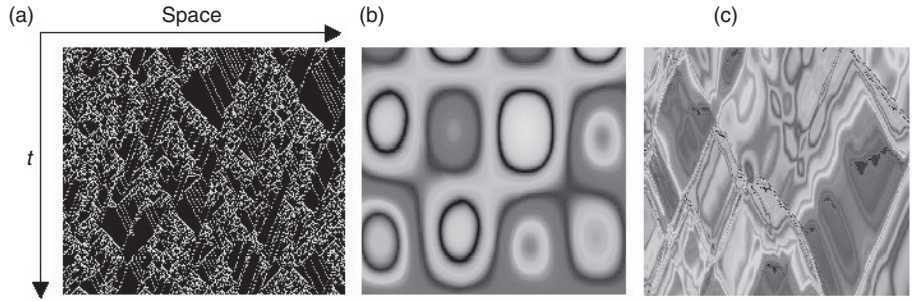

(a) Space    (b)    (c)

$t$

**FIGURE 18.3**  Pattern formation in cellular automata: (a) 1D CA with disordered initial conditions; (b) CA for 1D wave equation; and (c) nonlinear 1D Sine-Gordon equation.

of the typical characteristics in cellular automata. The study of pattern complexity and the conditions of its formation is essential in many processes such as biological pattern formation, enzyme dynamics, percolation, and other processes in engineering applications. This section focuses on the pattern formation in cellular automata and comparison of CA results with the results using differential equations.

### 18.5.1  Complexity and Pattern in Cellular Automata

Wolfram (1983) pioneered the studies of complexity and patterns in cellular automata. Complex patterns form in 1D cellular automata even with very simple rules. Figure 18.3(a) shows the typical patterns in 1D cellular automata with random initial conditions for $k = 16$ states and $r = 1$. Figure 18.3(b) is for 1D wave equation with $k = 1024$ states and $r = 1$. Figure 18.3(c) corresponds to the nonlinear wave based on Sine-Gordon equation $u_{tt} - u_{xx} = \alpha \sin(u)$ with $\alpha = -0.1$, $k = 1024$ states, $r = 2$ and sinusoidal initial conditions. The pattern formations are more complicated in higher dimensions. We compare the results from different methods while we study the pattern formations in the next section.

### 18.5.2  Comparison of Cellular Automata and PDEs

For a 2D reaction-diffusion system of two partial differential equations:

$$\frac{\partial u}{\partial t} = D_u \nabla^2 u + f(u), \quad \frac{\partial v}{\partial t} = D_v \nabla^2 v + g(v), \quad f(u, v) = \alpha(1 - u) - uv^2,$$

$$g(u, v) = uv^2 - \frac{(\alpha + \beta)v}{1 + (u + v)},$$

where $D_u = 0.05$. The parameters $\gamma = D_v/D_u$, $\alpha$, $\beta$ can vary so as to produce the complex patterns. This system can model many systems such as enzyme dynamics and biological pattern formations by slight modifications (Murray, 1989; Meinhardt, 1995; Keener and Sneyd, 1998; Yang, 2003). Figure 18.4 shows a snapshot of patterns formed by the simulations of the above reaction-diffusion system at $t = 500$ for $\gamma = 0.6$, $\alpha = 0.01$, and $\beta = 0.02$. The right plot is the comparison of results obtained by three different methods: cellular automata (marked with CA), finite difference method (FD), and finite element method (FE). The plot is for the data on the middle line of the pattern shown on the left.

### 18.5.3  Pattern Formation in Biology and Engineering

Pattern formation occurs in the many processes in biology and engineering. We will give two examples here to show the complexity and diversity of the beautiful patterns formed. Figure 18.5 shows the 2D and 3D patterns for the reaction diffusion system with $f(u, v) = \alpha(1 - u) - uv^2/(1 + u + v)$, $g(u, v) = uv^2 - (\alpha + \beta)v/(1 + u + v)$.

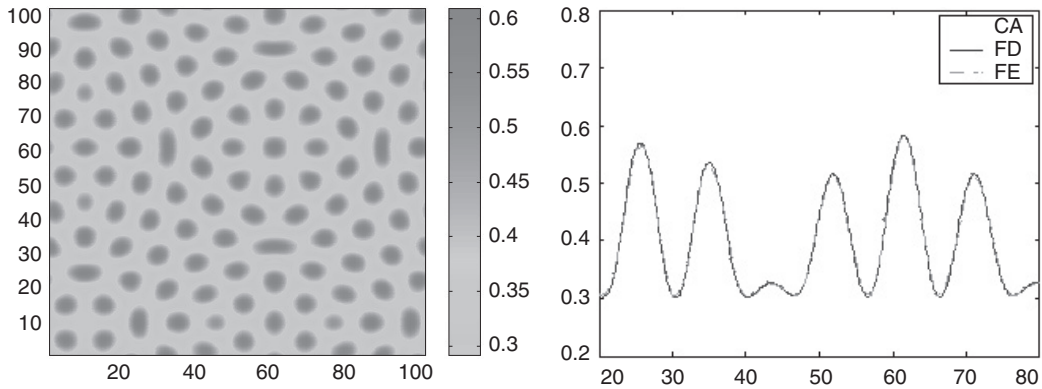

**FIGURE 18.4**  A snapshot of pattern formation of the reaction-diffusion system (for $\alpha = 0.01$, $\beta = 0.02$, and $\gamma = 0.6$) and the comparison of results obtained by three different methods (CA, FD, FE) through the middle line of the pattern on the left.

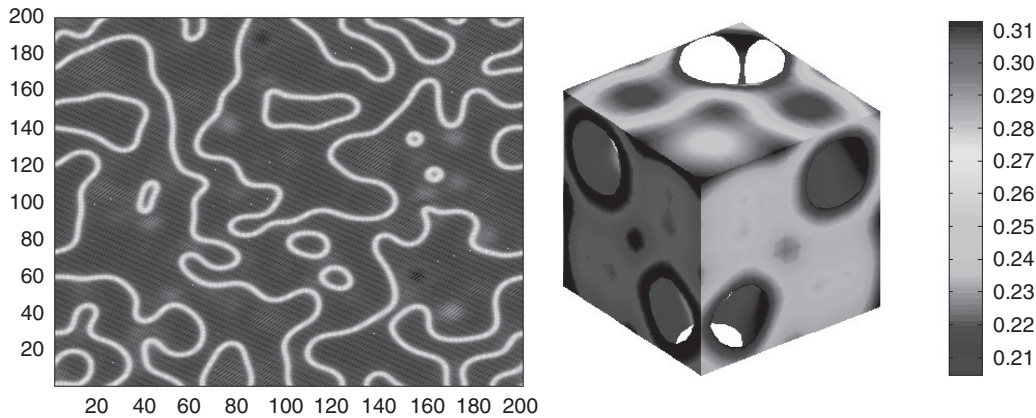

**FIGURE 18.5**  Pattern formations from 2D random initial conditions with $\alpha = 0.05$, $\beta = 0.01$, and $\gamma = 0.5$ (left) and 3D structures with $\alpha = 0.1$, $\beta = 0.05$, and $\gamma = 0.36$ (right).

Another example is the formation of spiral waves studied by Barkley and his colleagues in detail (Barkley et al., 1990; Margerit and Barkley, 2001)

$$\frac{\partial u}{\partial t} = \nabla^2 u + \frac{u}{\varepsilon^2}(1-u)\left(u - \frac{v+\beta}{\alpha}\right), \qquad \frac{\partial v}{\partial t} = \varepsilon \nabla^2 v + (u - v),$$

where $\varepsilon, \alpha, \beta$ are parameters. Figure 18.6 shows the formation of spiral waves (2D) and scroll waves (3D) of this nonlinear system under appropriate conditions.

The patterns formed in terms of diffusion-reaction equations have been observed in many phenomena. The ring, spots, and stripes exist in animal skin coating, enzymatic reactions, shell structures, and mineral formation. The spiral and scroll waves and spatiotemporal pattern formations are observed in calcium transport, Belousov–Zhabotinsky reaction, cardiac tissue, and other excitable systems.

In this chapter, we have discussed some of important development and research results concerning the connection among the cellular automata and partial differential equations as well as the pattern formation related to both systems. Cellular automata are rule-based methods with the advantages of local interactions, homogeneity, discrete states and parallelism, and thus they are suitable for simulating systems with large degrees of freedom and life-related phenomena such as artificial intelligence and ecosystems.

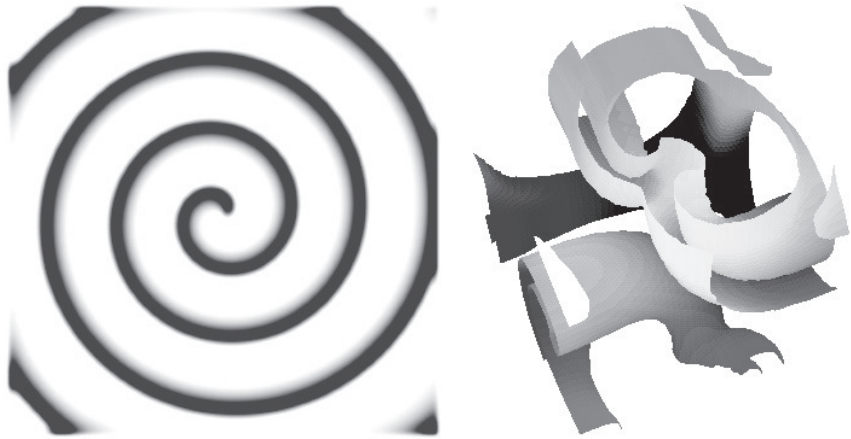

**FIGURE 18.6**    Formation of spiral wave for $\alpha = 1$, $\beta = 0.1$, and $\varepsilon = 0.2$ (left) and 3D scroll wave for $\alpha = 1$, $\beta = 0.1$, and $\varepsilon = 0.15$ (right).

PDEs are continuum-based models with the advantages of mathematical methods and closed-form analytical solutions developed over many years, however, they usually deal with systems with small numbers of degree of freedom. There is an interesting connection between cellular automata and PDEs although this is not always straightforward and sometimes may be very difficult. The derivation of updating rules for cellular automata from corresponding PDEs are relatively straightforward by using the finite differencing schemes, while the formulation of differential equations from the cellular automaton is usually difficult and nonunique. More studies are highly needed in these areas. In addition, either rule-based systems or equation-based systems can have complex pattern formation under appropriate conditions, and these spatiotemporal patterns can simulate many phenomena in engineering and biological applications.

AQ: Please provide citation for Boffetta et al. (2002); Cappuccio et al. (2001); Flake (1998); Meinhardt (1982); Turing (1952); von Newman (1966) & Ostrov and Rucker (1996)

AQ: Year has been introduced from the text for Rucker, R. et al. Please check.

AQ: Please provide the chapter title for von Newman (1966).